\documentclass[preprint,groupedaddress,a4paper,floatfix,aps,pre]{revtex4-1}
\usepackage[dvips]{graphicx}
\newcommand{\dif}{\mathrm{d}}
\newcommand{\del}{\partial}

\begin{document}

\begin{center}
\includegraphics[scale=0.6]{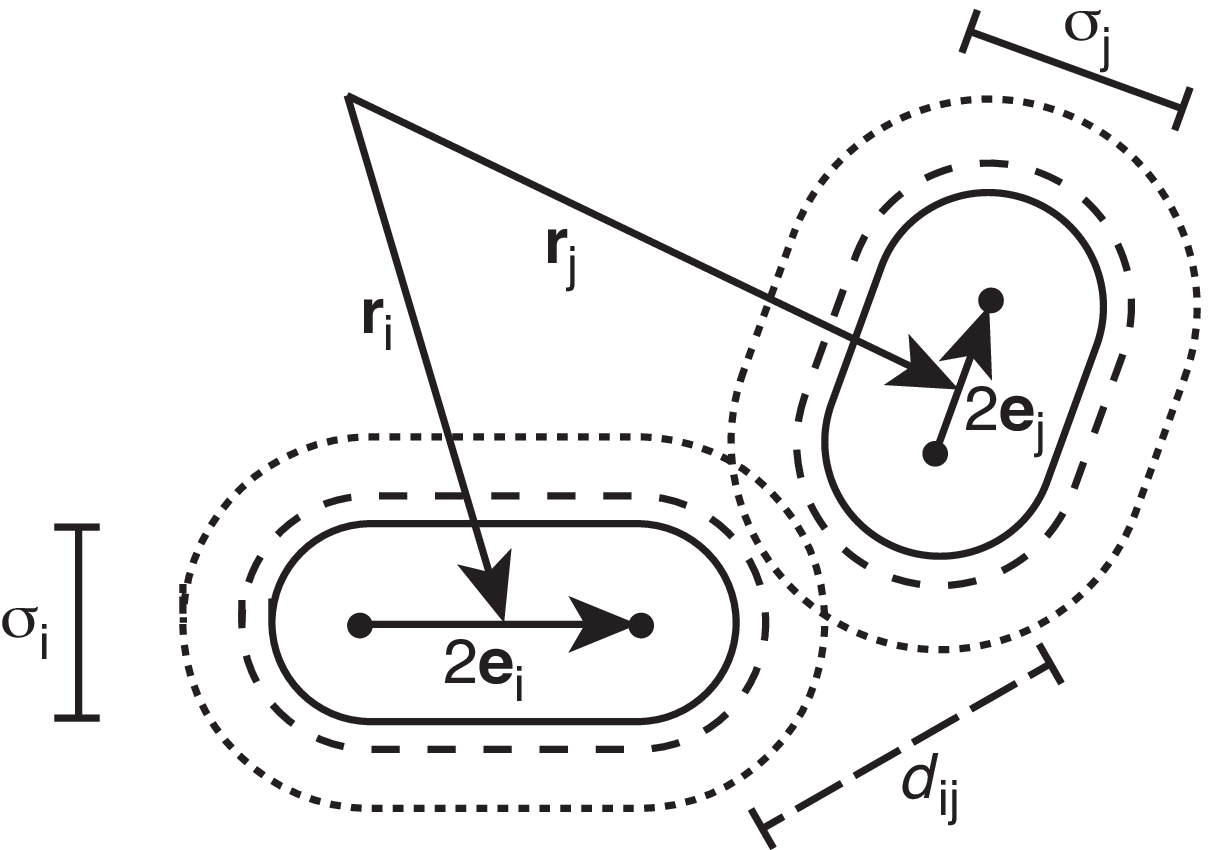}
\end{center}
For TOC use only: A new simulation model can elucidate the relation between rheology
and the properties of latex particles forming a pressure sensitive adhesive. Each latex
particle is treated as a single extensible particle. Transient forces are 
essential to correctly predict qualitative features.

\newpage

\title{Mesoscale modeling of the rheology of pressure sensitive adhesives through inclusion of transient forces}

\author{J.~T. Padding$^1$\footnote{Corresponding author, e-mail: \texttt{j.t.padding@gmail.com}}}
\author{L.~V. Mohite$^1$}
\author{D.~Auhl$^1$}
\author{W.~J. Briels$^2$}
\author{C. Bailly$^1$}
\affiliation{$^1$ Institut de la Mati\`ere Condens\'ee et des Nanosciences, Universit\'e catholique de Louvain, Croix du Sud 1, 1348 Louvain-la-Neuve, Belgium \\
$^2$ Computational Biophysics, University of Twente, PO Box 217, 7500 AE, Enschede, The Netherlands}

\date{\today}

\begin{abstract}
For optimal application, pressure-sensitive adhesives must have rheological properties in between those of a viscoplastic solid
and those of a viscoelastic liquid. Such adhesives can be produced by emulsion polymerisation, resulting in latex particles which are
dispersed in water and contain long-chain acrylic polymers. When the emulsion is dried, the latex particles coalesce and an
adhesive film is formed. The rheological properties of the dried samples are believed to be dominated by the interface regions
between the original latex particles, but the relation between rheology and latex particle properties is poorly understood.
In this paper we show that it is possible to describe the bulk rheology of a pressure-sensitive adhesive by
means of a mesoscale simulation model. To reach experimental time and length scales, each latex particle is represented by
just one simulated particle. The model is subjected to oscillatory shear flow and extensional
flow. Simple order of magnitude estimates of the model parameters already lead to semi-quantitative agreement with
experimental results. We show that inclusion of transient forces in the model, i.e. forces with memory of previous
configurations, is essential to correctly predict the linear and nonlinear properties.

\end{abstract}





\maketitle

\section{Introduction}

Pressure sensitive adhesives (PSAs) are tacky viscoelastic or viscoplastic materials that adhere to a substrate upon the
application of light pressure \cite{Creton2003}. A well known example of a PSA is the sticky layer on Scotch\textregistered \ tape.
Many modern PSAs are soft nanostructured materials formed by drying of an aqueous dispersion of polymeric
particles, also known as a latex. Such polymeric adhesives offer many advantages over rubber-based or ceramic alternatives, like
for example a low plateau modulus, superior stability, great resistance to oxidation, and avoidance of harmful or volatile organic solvents. 
Unfortunately, the design of new PSAs is hampered by a lack of fundamental understanding of the link between microscopic parameters,
such as polymer molecular weight and strength of polar interactions, and the macroscopic rheology. Computer simulations
can help in increasing this understanding, but no particle-based models are available that can reach large enough time and
length scales to predict the experimentally accessible rheology.

In this paper we introduce a novel highly coarse-grained simulation model
that can predict semi-quantitatively the bulk rheology of a PSA. We will show that inclusion of transient forces \cite{Briels},
i.e. forces with memory of previous configurations, is essential to correctly predict the linear and nonlinear properties.
Without transient forces even the qualitative features are incorrect.

A large fraction of industrial waterborne PSAs is based on acrylic polymers synthesised via radical emulsion
polymerisation processes \cite{Lovell}. Usually the formulation is complex, employing random copolymers of a long-chain
acrylic (such as $n$-butyl acrylate) characterised by a low glass transition temperature ($T_g$) with a short side-chain acrylic
(such as methyl acrylate) to adjust $T_g$ and acrylic acid to improve adhesion and optimise elongational properties \cite{Amaral,Lindner}.
When the emulsion is dried, the latex particles coalesce and an adhesive film is formed.

\begin{figure}[tbp]
\begin{center}
\includegraphics[scale=0.7]{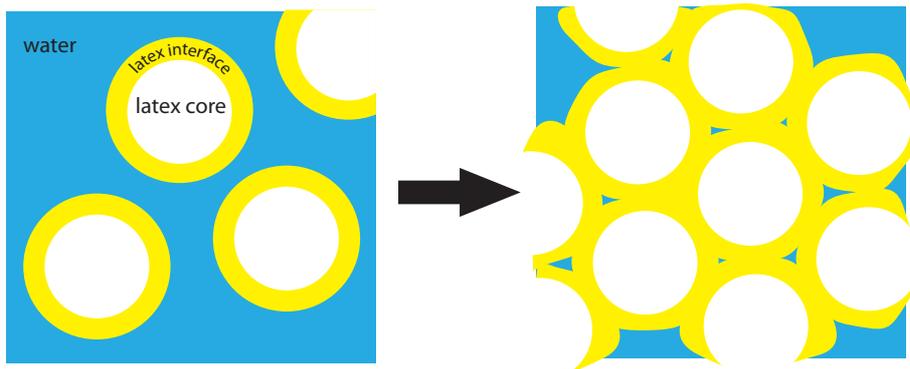}
\end{center}
\caption{\label{fig_cartoon}
(color online) 
Cartoon of the creation of an adhesive film. Left: polymeric latex particles (white) are dispersed in water (blue) by the action of surfactants and
polar groups which predominantly reside on and in the interfacial region of the latex particles (yellow). Right: when the system is dried, the water 
evaporates and most, but probably not all, surfactants are drained away. The polymer chains from different latex particles start to intermix,
and polar groups cause a strong bonding between the particles.
}
\end{figure}

At first sight the adhesive film appears to be homogenous. However there are indications that a fundamental heterogeneity
is retained within the structure, which is linked to an incomplete coalescence of the latex particles \cite{Mallegol,Mallegol2}, see Fig.~\ref{fig_cartoon}.
Memory of the original interface between latex particles could be retained for a number of reasons: (i) because surfactants, used
during the emulsion polymerisation process and used to stabilise the latex dispersion, do not drain away fully but partly get stuck in the
interfacial regions \cite{Arnold},
(ii) because polymer chains on different latex particles do not have enough time to fully interdiffuse, or
(iii) because polar groups (such as groups of acrylic acid and sodium ions) preferably reside in the outer regions of the original latex particles
and do not homogenise after coalescence. A combination of these effects is also possible.
Heterogeneities in the concentration of remaining surfactants can be visualised by transmission electron
microscopy, and heterogeneities in polymer and polar group concentrations can be characterised by atomic force microscopy \cite{Mallegol,Mallegol2,Pinprayoon,Canetta}.
The effects of heterogeneities can also be traced back in the dynamical behaviour of the film.
The bulk rheology for example is more complex than that of a homogeneous
melt of polymers with comparable molecular weight and branching frequency.

Not surprisingly, the design of new waterborne PSAs is a complex formulation process, in which the number of parameters is so large
that the approach is often empirical and far from optimised. To enable a knowledge-based design, a fundamental understanding is needed
of the link between the molecular properties and structure of the latex particles on the one side, and the mechanical properties of the
adhesive film on the other side. Particle based computer simulations could provide such a link.

Here we present a novel mesoscopic Brownian Dynamics (BD) simulation model in which the smallest unit is one full
latex particle produced by emulsion polymerisation.
This degree of coarse graining is as large as possible to permit a large integration step (3 10$^{-5}$ s) and
few degrees of freedom. The key issue is to identify the most important microscopic properties of a PSA that
determine its rheology. In our work we have assumed that these key properties are the
mechanical properties that characterise a single latex particle, such as its size and deformability, 
and the properties of the interfacial region, in particular the interaction
free energies and relaxation times associated with latex interparticle adhesion and polymer chain intermixing. 
It is important to note that in our one-particle approach the latex particles keep their individuality,
even after drying. It is therefore implicitly assumed that no major mass transfer between the latex particles will occur.
Whether or not this is the case for real latex particles is presently unclear, but the observed heterogeneities
suggest that interdiffusion is probably not dominant.

Before embarking on a programme in which the model is tweaked and optimised to quantitatively predict the rheology of specific
PSAs, our first goal is to test whether our model can predict the qualitative features.
We found that using crude estimates for the model parameters already enabled us to reproduce 
semi-quantitatively the linear rheology of a particular experimental PSA, which we refer to as our ``reference material''. 
We will subsequently compare \textit{predictions} for the non-linear extensional rheology with experimental measurements.
As already mentioned, inclusion of memory effects, through application of transient forces, is necessary to correctly predict
the linear and nonlinear dynamic properties.

This paper is organised as follows. 
The BD model, including the treatment of transient forces, is explained in section \ref{sec_model}.
Details of the reference material, its preparation and, coupled to this, a justification for the simulation parameters, are given in section \ref{sec_material}.
In section \ref{sec_exp} we describe the rheological experiments we have performed on the reference material.
In sections \ref{sec_linear} and \ref{sec_nonlinear} we present
the simulation predictions for the linear and nonlinear rheology, respectively, and compare with experimental measurements 
on the reference system.
We end with conclusions and an outlook on further developments of the model in section \ref{sec_concl}.

\section{Mesoscale model with transient forces\label{sec_model}}

\subsection{Coarse graining}

Our goal is to describe the configurations and forces in a pressure sensitive adhesive formed by drying of a latex emulsion
by as few variables as possible, while at the same time retaining a link with the chemistry and individuality of the latex particles.
We choose to represent the configuration by just six coordinates for each latex particle, corresponding to a vector locating its
centre of mass position and a vector characterising its extension.
This is not to say that all the removed coordinates are irrelevant for the rheology of the system. They provide an effective potential,
a so-called potential of mean force $A^{mf}$, governing the equilibrium distribution of the centres of mass and extensions.
In thermodynamic equilibrium the probability
distribution $P_{eq}\left(r^{3N},e^{3N}\right)$ of the particles' centre-of-mass positions and extension vectors is given by
\begin{equation}
P_{eq}\left(r^{3N},e^{3N}\right) \propto \exp\left[-\frac{A^{mf}\left(r^{3N},e^{3N}\right)}{k_BT}\right].
\end{equation}

It is possible to derive equations of motion for the retained coordinates, not only leading to a mean force (which is minus the gradient
of the potential of mean force), but also to frictions and random forces \cite{DeutchOppenheim,MazurOppenheim}.
In most coarse-grained models of soft matter systems, these frictions and random
forces have memory of the configurations the system has gone through in the recent and sometimes even the distant past \cite{Briels}.
In such cases a simple Brownian dynamics propagator with realistic mean forces and uncorrelated, fully random displacements without
memory will not reproduce correct sequences of configurations of the retained coordinates. To circumvent the introduction
of memory effects in friction forces and stochastic displacements we employ the Responsive Particle Dynamics (RaPiD) method \cite{vdNoort1,vdNoort2,Kindt,vdNoort3,Briels,Sprakel,Padding}.

The idea behind the RaPiD method is to introduce a relatively small set of additional dynamic variables which keep
track of the thermodynamic state of the eliminated coordinates for the given values of the retained coordinates. The deviations of these
additional variables from their equilibrium values at the given configuration give rise to strong forces in addition to the
thermodynamic forces deriving from the potential of mean force. While the additional non-equilibrium forces gradually fade away because
the variables relax towards their equilibrium values, with every new time step new such forces are being generated because the displaced
coordinates dictate new equilibrium values for the variables. For this reason the additional forces are called transient forces.

In the following subsections we will introduce our model for the potential of mean force and the transient forces. 
We will give the equations of motion by which we update the retained coordinates and additional variables characterising deviations
from equilibrium.

\subsection{Potential of mean force}
\begin{figure}[tbp]
\begin{center}
\includegraphics[scale=0.6]{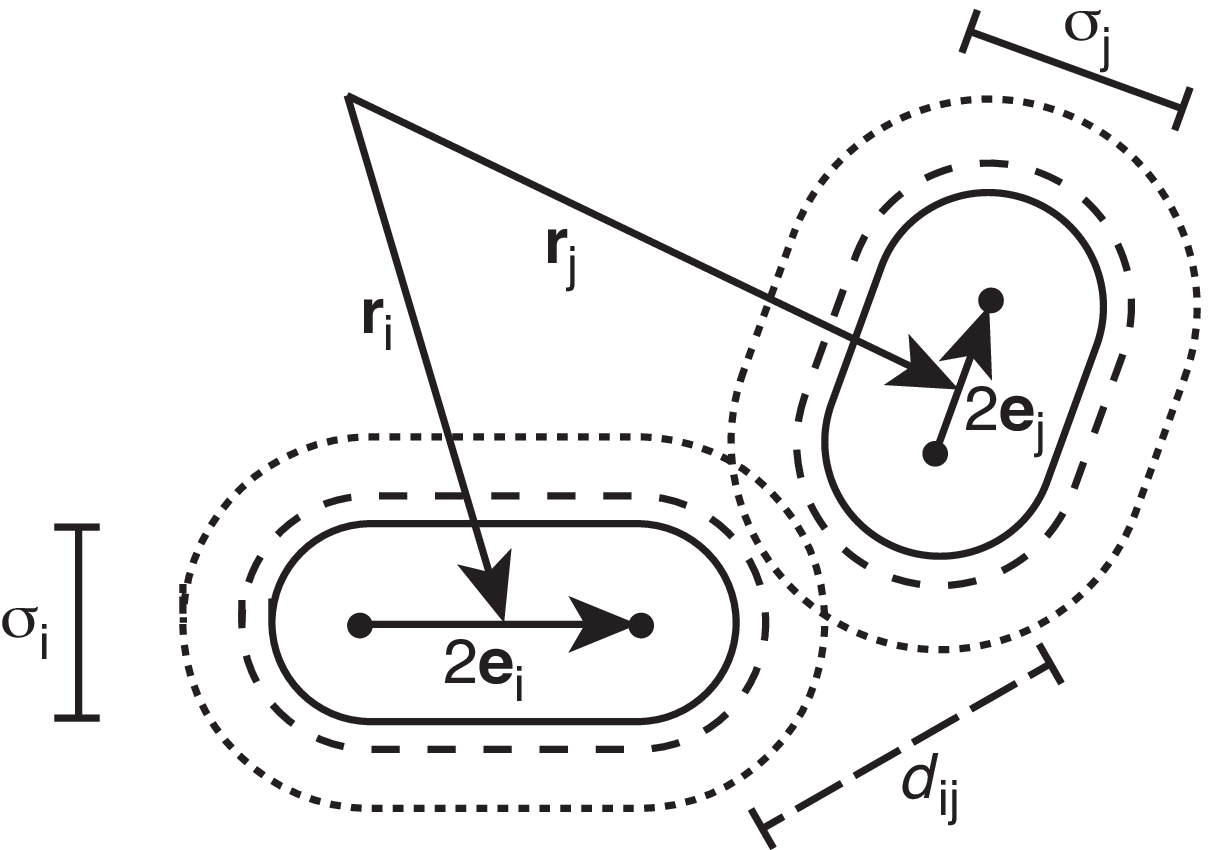}
\end{center}
\caption{\label{fig_geometry}
Each latex particle is represented by a hemispherical cylinder located at $\mathbf{r}$, 
with an orientation and extension given by $\mathbf{e}$ (cylinder length 2e).
A finitely extensible nonlinear elastic spring is associated with each extension.
All interactions between pairs of particles $i$ and $j$ are functions of
the closest distance $d_{ij}$ between the line segments running through the cylinder axes.
The diameter $\sigma_i$ of a cylinder decreases with increasing extension $e_i$ in such a way
that the volume of the hemispherical cylinder is conserved.
}
\end{figure}

The latex particles are relatively soft and deformable. In the simulation we allow the latex particles to deform and orient
by representing each particle, labeled $i$, as a hemispherical cylinder with an internal finitely extensible nonlinear elastic (\textit{fene})
spring of length $2e_i$, as shown in Fig.~\ref{fig_geometry}.
The diameter $\sigma_i$ of the cylinder is defined as $\sigma_0$ at zero extension (when the particle is actually a sphere). With increasing
extension the diameter decreases such that the volume of the hemispherical cylinder remains equal to  $\frac16 \pi \sigma_0^3$. An analytical function for the diameter may be derived by solving a cubic equation using Cardano's rules, but in practice in the range of extensions sampled in this 
work ($0 \leq 2e_i < 2.5\sigma_0$) a good fit is given by the much simpler function
\begin{equation}
\sigma_i(e_i) = \sigma_0 \left[1+a_1\left( \frac{e_i}{\sigma_0} \right) \right]^{a_2},
\end{equation}
with $a_1 = 1.59$ and $a_2 = -0.61$.

Figure \ref{fig_geometry} also shows how we let the interaction between each pair of particles be a function of the closest distance
$d_{ij}$ between the line segments running through the cylinder axes:
\begin{equation}
d_{ij} = \min_{s_i,s_j \in [-1,1]} \left\{ \left| \mathbf{r}_i + s_i \mathbf{e}_i - \mathbf{r}_j -
s_j \mathbf{e}_j \right| \right\} \label{eq_dij}
\end{equation}
Here $\mathbf{r}_i$ and $\mathbf{e}_i$ are the location of the centre and (half) the extension of dumbbell $i$, respectively.
Note that the values of $s_i$ and $s_j$ are constrained to the interval $[-1,1]$.
In practice, we first solve analytically the unconstrained minimisation of the bivariate quadratic function
$d_{ij}^2(s_i,s_j)$. If the resulting values of $s_i$ and $s_j$ are not inside the square 
represented by $s_i = \pm 1$ and $s_j = \pm 1$, then each of the sides of this square are checked.
The latter corresponds to a situation in which an end-point of the segment
is in closest contact with the other line segment.

We let the particles interact with a repulsive interaction $\varphi^{rep}$ and an attractive interaction $\varphi^s$,
both of which scale with the average particle diameter $\sigma_{ij} = \frac12 (\sigma_i + \sigma_j)$.
Physically, the repulsive interaction $\varphi^{rep} = \varphi^{rep}(d_{ij}/\sigma_{ij})$ results from the entropic cost of chain deformation,
as well as more direct excluded volume interactions between monomers. 
The former sort of repulsion is relatively soft and important at intermediate distances between latex particles,
whereas the latter is stronger and more important at close distances. The latex particles do not only repel each other, but also attract because 
the polar groups generate strong bonds between individual chains.
In the model we assume that the free energy associated with the cohesion between a pair of latex particles changes from $-\epsilon^s$ to zero with increasing scaled interparticle distance through a smooth `sticker' potential function $\varphi^s = \varphi^s(d_{ij}/\sigma_{ij})$.

In detail, the total potential of mean force has the following form:
\begin{eqnarray}
A^{mf} &=& \sum_i \varphi^{fene}(e_i) + \sum_{i<j} \left\{ \varphi^{rep}(d_{ij},\sigma_{ij}) + \varphi^s(d_{ij},\sigma_{ij}) \right\} \\
\varphi^{fene}(e) &=& -\frac12 k R_0^2 \ln \left[ 1 - \left( \frac{2e}{R_0} \right)^2 \right] \label{eq_fene} \\
\varphi^{rep}(d,\sigma) &=& 4a^{rep} \left[ \left( \frac{\sigma}{d} \right)^{2n} - \left( \frac{\sigma}{d} \right)^n + \frac14 \right] \qquad (d < 2^{1/n} \sigma) \\
\varphi^s(d,\sigma) &=& -\epsilon^s n_0^s(d,\sigma) \\
n_0^s(d,\sigma) &=& \frac12 \left[ 1 - \tanh \left( \frac{\sigma_0}{w^s} \left( \frac{d}{\sigma} - 1 \right) -1 \right) \right] \label{eq_n0s}
\end{eqnarray}
\begin{figure}[tbp]
\begin{center}
\includegraphics[scale=0.5]{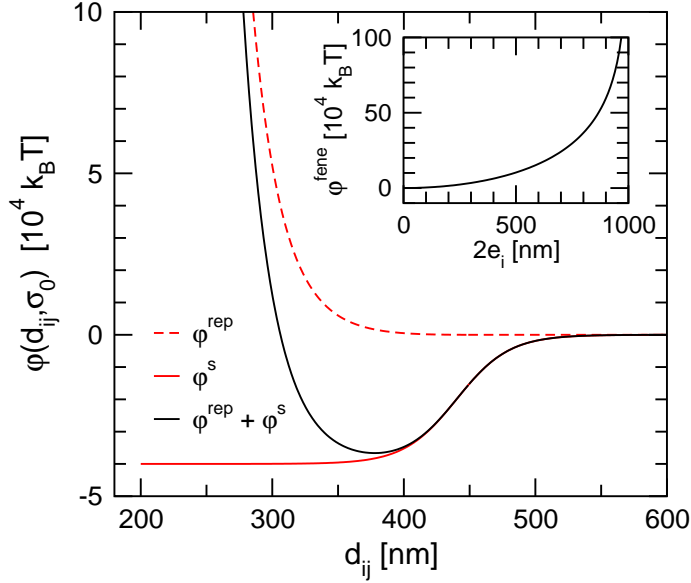}
\end{center}
\caption{\label{fig_pmf}
(color online)
Potential of mean force, with the parameters used in this work.
Main figure: repulsive (dashed red line) and attractive sticker (solid red line) pair interaction as a function of 
distance $d_{ij}$ for $\sigma_{ij}=\sigma_0 = 400$ nm. The total pair interaction (solid black line) has an attractive well around $\sigma_0$.
Inset: potential of the finitely extensible nonlinear elastic spring as a function of the total intraparticle extension $2e_i$. The maximum
extension in this work is $R_0 = 1000$ nm.
}
\end{figure}
The fene potential in Eq.~(\ref{eq_fene}) behaves as a linear spring with spring constant $k$ at small extensions, while it diverges asymptotically
at the maximum extention $R_0$. The repulsive potential is characterised by its softness parameter $n$ and equal to
$a^{rep}$ at $d=\sigma$. For low values of $n$ the repulsive interaction is, as desired, relatively soft in the range
$\sigma < d < 2^{1/n} \sigma$ and quickly becomes stronger for $d < \sigma$, where it diverges as $r^{-2n}$. 
The particles can therefore interpenetrate, but not so far that they can
fully 'cross each other'.
The repulsive, sticker and fene interactions are shown in Fig.~\ref{fig_pmf}.

Note that we have chosen, rather arbitrarily, a tangenthyperbolic function for the sticker potential of mean force.
The corresponding function $n_0^s(d,\sigma)$ may be thought of as the average fraction of sticker groups still active between
a pair of particles at distance $d$. The only requirement for this function is that it is
smooth and decays from one to zero over a finite width $w^s$. 
We have chosen its inflection point at $\sigma_0 + w^s$ (scaled by $\sigma/\sigma_0$), which ensures that almost all sticker
groups are active when $d \leq \sigma$.

We will give estimates for the parameters appearing in the potential of mean force in subsection
\ref{subsect_parameters}. In principle the potential of mean force can also be obtained directly
by means of an atomic force microscope by measuring the average forces between two latex particles held at various distances, although in practice this has not been done yet.

\subsection{Transient forces}

For the current system we identify two major sources of transient forces, namely deviations in the number of sticker groups and
deviations in the amount of `intermixing' of the chains. The latter is a
topological effect: when the distance between two latex particles is suddenly changed,
the relaxation from unfavorable polymer configurations to new favorable ones
is slowed down by the connectivity and uncrossability of the polymer chains.
We denote the fraction of intermixing of the chains belonging to particles $i$ and $j$
by the symbol $n^e_{ij}$, where the `e' stands for entanglement, with the understanding
that any topological effect is included: these are not literally the entanglements treated in the 
Doi and Edwards tube model \cite{DoiEdwards}, but rather a measure for intermixing between the chains belonging to different latex particles.
Similarly, the symbol $n^s_{ij}$ is used for the actual fraction of sticker groups 
shared between latex particles $i$ and $j$. We assume that the
equilibrium number of sticker groups and the equilibrium amount of chain mixing 
between $i$ and $j$ are functions only of the distance between these two particles. 

It is possible to show \cite{Briels} that, to lowest order, the transient forces may be
thought of as originating from a penalty free energy function $A^{trans}$ which is quadratic in the deviations of the additional variables
from their equilibrium values, i.e.
\begin{equation}
A^{trans} = \sum_{i<j} \left\{ \frac12 \alpha^s \left[ n_{ij}^s - n_0^s\left(d_{ij},\sigma_{ij}\right) \right]^2
+ \frac12 \alpha^e \left[ n_{ij}^e - n_0^e\left(d_{ij},\sigma_{ij}\right) \right]^2 \right\},
\end{equation}
where, similar to the average number of sticker groups, the average amount of chain mixing
between a pair of latex particles is taken to be a smooth general S-shaped function:
\begin{equation} 
n_0^e(d,\sigma) = \frac12  \left[ 1 - \tanh \left( \frac{\sigma_0}{w^e} \left( \frac{d}{\sigma}-1 \right) \right) \right].
\end{equation}
Here $w^e$ gives the characteristic width over which the chain mixing decays smoothly from one for $d < \sigma$ to zero for 
$d > \sigma$.

The new variables $\alpha^s$ and $\alpha^e$ control the allowed equilibrium deviations in the number of sticker groups and chain mixing.
Generally, larger values of $\alpha^s$ and $\alpha^e$ correspond to larger penalties for deviations, resulting in stronger transient forces. 

\subsection{Equations of motion for the retained coordinates}

In the preceding subsections we have defined the potential of mean force $A^{mf}$ and the penalty free energy $A^{trans}$ 
associated with transient forces.
The actual simulation proceeds by integrating (by a simple first order Euler scheme) the following stochastic equations of motion:
\begin{eqnarray}
\frac{\dif \mathbf{r}_i}{\dif t} &=& - \frac{1}{\xi^r_i} \frac{\del A}{\del \mathbf{r}_i} + \mathbf{f}^r(t) \label{eq_dr} \\
\frac{\dif \mathbf{e}_i}{\dif t} &=& - \frac{1}{\xi^e_i} \frac{\del A}{\del \mathbf{e}_i} + \mathbf{f}^e(t) \label{eq_de}
\end{eqnarray}
Here $A = A^{mf} + A^{trans}$ is the total free energy of the system and $\xi^r$ and $\xi^e$ are frictions on displacements of
the centre of mass and internal deformations, respectively. We need a model to describe these friction functions. We will assume that
the frictions depend linearly on the amount of chain mixing with neighbouring particles.
In particular, we use:
\begin{equation}
\xi^r_i = \xi^e_i = \xi_e \sum_{j \ne i} \sqrt{n^e_{ij}n^e_0(d_{ij})}, \label{eq_xi}
\end{equation}
were $\xi_e$ is the maximum friction generated by the mixed interface between a pair of particles.
For simplicity, we have made the frictions on the positions and the extensions equal. Although this can be relaxed
in future work, it is expected that the two frictions will not be very different because 
a change in position $\mathbf{r}_i$ generates roughly the same amount of relative motion
as a similar change in extension $\mathbf{e}_i$.
In Eq.~(\ref{eq_xi}) we use a form $\sum_j \sqrt{n^e_{ij}n^e_0(d)}$, 
rather than $\sum_j n^e_{ij}$, because this allows the friction contributions to go to zero smoothly at
the cut-off distance. In practice we use large values of $\alpha^e$ and therefore $n^e_{ij}$ will always be close to $n^e_0$, and the
difference will therefore be small.

In principle, to ensure a correct equilibrium probability density, also a `drift' term $\del/\del \mathbf{r}_i (k_BT/\xi^r_i)$ 
should be added to Eq.~(\ref{eq_dr}), and a similar
term to Eq.~(\ref{eq_de}), but in our simulations the densities are always so high that the gradients in friction are negligible compared
with gradients in the free energy.

Finally, the random forces $\mathbf{f}^r(t)$ and $\mathbf{f}^e(t)$ are related to the frictions through the fluctuation-dissipation theorem,
resulting in:
\begin{eqnarray}
\left\langle \mathbf{f}^r (t) \mathbf{f}^r (t') \right\rangle &=& \frac{2k_BT}{\xi^r_i}\mathbf{I} \delta(t-t') \\
\left\langle \mathbf{f}^e (t) \mathbf{f}^e (t') \right\rangle &=& \frac{2k_BT}{\xi^r_e}\mathbf{I} \delta(t-t')
\end{eqnarray}
Here $\mathbf{I}$ is the unit tensor and $\delta (t)$ is the Dirac delta function.

\subsection{Equations of motion for the sticker and entanglement numbers}

Besides the equations of motion for the retained coordinates, we also need equations of motion for the sticker and chain mixing fractions,
$n^s_{ij}$ and $n^e_{ij}$.
These numbers tend to relax towards the equilibrium values prevailing at the given particle configuration.	
It is important to realise that such a relaxation is not instantaneous, but requires a certain amount of time.
This is exactly what gives rise to memory and transient forces.

We update the sticker and intermixing fractions according to: 
\begin{eqnarray}
\frac{\dif n^s_{ij}}{\dif t} &=& -\frac{1}{\tau^s} \left[ n^s_{ij} - n^s_0\left( d_{ij} \right) \right] + g^s(t) \\
\frac{\dif n^e_{ij}}{\dif t} &=& -\frac{1}{\tau^e} \left[ n^e_{ij} - n^e_0\left( d_{ij} \right) \right] + g^e(t) \\
\left\langle g^s (t) g^s (t') \right\rangle &=& \frac{2k_BT}{\alpha^s \tau^s}\delta(t-t') \label{eq_amplgs} \\
\left\langle g^e (t) g^e (t') \right\rangle &=& \frac{2k_BT}{\alpha^e \tau^e}\delta(t-t') \label{eq_amplge}
\end{eqnarray}
Here $\tau^s$ and $\tau^e$ are the characteristic times with which deviations in the fraction of sticker groups and
chain mixing, respectively, relax towards their equilibrium values. 
For simplicity, in this work the two characteristic times are (different) constants. It is likely that the relaxation times actually
depend on the distance between a pair of particles (i.e. their overlap). This may be included in future work
when the model is refined further.


The amplitudes of the random terms $g^s$ and $g^e$ in Eqs.~(\ref{eq_amplgs}) and (\ref{eq_amplge})
are chosen such that in equilibrium the expected fluctuations in sticker and mixing fractions are given by:
\begin{eqnarray}
\left\langle \left[ n^s_{ij} - n^s_0\left( d_{ij} \right) \right]^2 \right\rangle &=& \frac{k_BT}{\alpha^s} \label{eq_flucs} \\
\left\langle \left[ n^e_{ij} - n^e_0\left( d_{ij} \right) \right]^2 \right\rangle &=& \frac{k_BT}{\alpha^e} \label{eq_fluce}
\end{eqnarray}
These latter two equations give us a convenient way to check (in equilibrium) whether the equations of motion are integrated
with a sufficiently small time step \cite{vdNoort1}: if the measured fluctuations in sticker and mixing fractions do not agree
with Eqs.~(\ref{eq_flucs}) and (\ref{eq_fluce}), a smaller time step must be used.

\subsection{Simulations without transient forces}

To assess the effect of transient forces, we also include simulations without transient forces. All conservative interactions,
i.e. all terms in the potential of mean force $A^{mf}$, are retained.  The transient penalty free energy $A^{trans}$, however, is set to
zero. For the friction we use exactly the same expression as before,
Eq.~(\ref{eq_xi}), but manually set $n_{ij}^e$ equal to the equilibrium number $n_0^e(d_{ij})$.

\subsection{Boundary conditions and nonlinear flow\label{sec_flowsim}}

As with other particle-based methods, our simulations are typically limited to systems containing several thousands of particles
(in this work 10,000). To mimic bulk behaviour, we apply periodic boundary conditions, where a primary box is surrounded by
copies of itself \cite{AllenTildesley}.
When a particle leaves the primary box, its image will simultaneously enter the box through the opposite face.
Therefore the shape of the box must be space filling. We choose a rectangular parallellipiped, with box sizes along $x$ and $y$
twice as large as the box size along $z$, $L_x = L_y = 2L_z$, for reasons that we will discuss next.

In this paper we apply two types of deformation, namely planar (oscillatory) shear flow and planar extensional flow.

Planar shear flow is characterised by a flow field
\begin{equation}
\mathbf{v}(\mathbf{r}) = \left(
\begin{array}{ccc}
0 & \dot{\gamma} & 0 \\
0 & 0 & 0 \\
0 & 0 & 0
\end{array}
\right) \cdot \mathbf{r}.\label{eq_shearflow}
\end{equation}
Such a flow field can be simulated by applying Lees-Edwards periodic boundary conditions \cite{AllenTildesley,LeesEdwards},
and augmenting the equations of motion with a flow given by Eq.~(\ref{eq_shearflow}).
The motion of a box image is such that its origin moves with a velocity in the $x$-direction which is proportional
to the $y$-coordinate of that particular box's origin. If a particle moves through the lower $y$-face of the primary box, the replacing image 
particle at the upper $y$-face will not have the same $x$-coordinate, but one displaced by an amount $\delta x = \dot{\gamma}L_y t$
(where $\delta x$ may be taken modulus $L_x$).

Planar extensional flow is characterised by a flow field
\begin{equation}
\mathbf{v}(\mathbf{r}) = \left(
\begin{array}{ccc}
\dot{\epsilon} & 0 & 0 \\
0 &-\dot{\epsilon} & 0 \\
0 & 0 & 0
\end{array}
\right) \cdot \mathbf{r}.
\end{equation}
It is more difficult to apply a continuous extensional flow to a periodic system. We use a method proposed by Todd and Daivis \cite{Todd1,Todd2}
and by Baranyai and Cummings \cite{Cummings}. Both methods are based on work of Kraynik and Reinelt \cite{Kraynik}.
The basic idea is to choose a primary box which is square in the $xy$-plane and rotated over some angle $\theta$. This angle is not arbitrary,
but must be chosen such that after a finite simulation time the deformed box coincides with some unit cell of the lattice generated by the original box. At this time all particles are mapped back into the original unit cell and the run is continued.
Out of the infinite set of solutions, we choose the case for which the minimum distance between opposite faces in the deformed box is still as
large as possible. Nevertheless, for this case the minimum distance is $\sqrt{5}$ times smaller than the original box size $L_x$ (or $L_y$) \cite{Todd1}. To prevent particles from interacting with their own images
it is important that the box size is always at least twice the maximum interaction range between a pair of particles \cite{AllenTildesley}.
This is the reason why the box dimensions in the $x$ and $y$ direction were chosen twice as large as in the $z$ direction. 

The response of the system to flow is measured by keeping track of the instantaneous microscopic stress tensor for a periodic system \cite{AllenTildesley}:
\begin{equation}
\mathbf{S} = - \frac{1}{V} \sum_{i < j} \left( \mathbf{r}_i - \mathbf{r}_j \right) \mathbf{F}_{ij}. \label{eq_micrstress}
\end{equation}
Here $V$ is the volume of the primary simulation box, 
$\mathbf{r}_i - \mathbf{r}_j$ is the vector pointing from the centre-of-mass of particle $j$ to the centre-of-mass of particle $i$, and
$\mathbf{F}_{ij}$ is the total interparticle force (both conservative and transient) exerted on particle $i$ by particle $j$. 
Because of the dominance of these interparticle interactions, we are justified in neglecting any kinetic contributions to the
microscopic stress. This is also implicit in our Brownian Dynamics approach.

\section{The reference material\label{sec_material}}

In this paper we aim to model the rheology of a particular pressure sensitive adhesive, the reference material.
In subsection \ref{subsect_materials}
we will briefly introduce the reference material and its preparation. In subsection \ref{subsect_parameters} we give an overview
of the parameters used in our simulations and a justification for their values.

\subsection{Materials and preparation\label{subsect_materials}}

The reference material consists mainly of acrylic polymers with a mean molecular weight of 2203 kg/mol. A latex emulsion with
53.6 \% solid content was received from DOW Chemical. An adhesive film was prepared by uniformly depositing about 2 ml of adhesive
emulsion into a flat rectangular silicone mold cavity ($23 \times 47 \times 6$ mm), which was then allowed to air dry at room temperature for
about one week. Finally, the now transparent films were dried in a vacuum oven for 5 minutes at 80$^{\circ}$C to remove remaining water.
These dried films were removed from the mold cavity and sandwiched between silicone papers in order to avoid sticking to any
other substrate during handling.

\subsection{Choice of simulation parameters for the reference material\label{subsect_parameters}}

An overview of the simulation parameters and the particular values used in this work is given in Table \ref{table_parameters}.
\begin{table}[tb]
	\centering
		\begin{tabular}{lcr}
		property & symbol & value \\
		\hline\hline
		General parameters \\
		\hline
		temperature & $T$ & 303 K \\
		volume fraction & $\phi$ & 1.0 \\
		\hline
		Particle \\
		\hline
		diameter & $\sigma_0$ & 400 nm \\
		softness parameter & $n$ & 6 \\
		repulsive energy (pair) & $a^{rep}$ & 500 $k_BT$ \\
		internal spring constant & $k$ & 3 10$^{-3}$ N/m \\
		maximum extension & $R_0$ & 1000 nm \\
		\hline
		Sticker clusters \\
		\hline
		adhesive energy (pair) & $\epsilon^s$ & 4\ 10$^4$ $k_BT$ \\
		transition width & $w^s$ & 40 nm \\
		transient penalty & $\alpha^s$ & 8 10$^5$ $k_BT$ \\
		relaxation time & $\tau^s$ & 1 s \\
		\hline
		Interparticle chain mixing \\
		\hline
		transition width & $w^e$ & 40 nm \\
		transitient penalty & $\alpha^e$ & 10$^5$ $k_BT$ \\
		relaxation time & $\tau^e$ & 10$^{-3}$ s \\
		friction per particle pair & $\xi_e$ & 2 10$^{-4}$ kg/s \\
		\hline
	         \end{tabular}
	\caption{Simulation parameters representing a reference pressure sensitive adhesive
	based on acrylic polymers. The interparticle chain mixing time $\tau^s$ and transient penalties $\alpha^s$ and
	$\alpha^e$ are fit parameters to semi-quantitatively tune the linear storage and loss modulus. All other parameters have been
	estimated from available data or statistical reasoning. }
	\label{table_parameters}
\end{table}
Most parameters of the model have been estimated from available data ($T, \phi, \sigma_0, w^e$ and $\tau^e$)
or statistical reasoning ($n, a^{rep}, k, R_0, \epsilon^s, w^s$ and $\xi^e$).
A few parameters, ($\tau^s$, $\alpha^s$ and $\alpha^e$)
could not be estimated and were treated as fit parameters to reproduce the experimental linear rheology. 

Transmission electron microscopy has revealed that before coalescence the latex particles have a diameter of $400$ nm.
After coalescence and drying of the film most solvent has evaporated. Assuming that the particles have not significantly swollen
or shrunk during the drying process, we have set our bare particle diameter to $\sigma_0 = 400$ nm. We have chosen 
the number density $\rho$ of latex particles such that the corresponding volume fraction $\phi = \frac{\pi}{6}\rho \sigma_0^3$ is
equal to 1.0.
For hard spheres such a high volume fraction would be unachievable ($\phi_{max} \approx 0.64$ for random close packing)
but it should be remembered that the latex particles have a polymeric content and are therefore soft and deformable.
This softness is reflected in a low value, $n=6$, for the scaling of the repulsive potential. We have also chosen a
relatively low value for the repulsive energy parameter $a^{rep}$ which is discussed after treating the sticker forces.
Together with the attractive sticker forces, the particles can interpenetrate quite easily (in contrast to hard spheres),
with closest distances as small as $d = 0.7 \sigma_0$ and a peak in the radial distribution function at $d = 0.85 \sigma_0$ (not shown).

The cohesion between the latex particles is mainly caused by polar charge interactions in clusters of acrylic acid -COOH or sulphate
-OSO$_3^-$ centered around sodium or other positively charged metal ions.
The binding energy for two carboxylates with identical sodium (in the absence of water) is of the order of 40 $k_BT$ \cite{Bendiksen,Remko}.
Each latex particle contains a very large number of polar charge groups on its surface. Based on the preparation chemistry,
and the fact that each latex particle on average interacts with 10 neighbouring latex particles,
we estimate that there are typically 1000 polar bonds between each pair of latex particles. This leads us to an estimate of 
$\epsilon^s = 4\ 10^4\ k_BT$ for the total adhesive energy between a pair of latex particles.

The characteristic width $w^s$ over which the attractive ``sticker'' potential $\varphi^s$ 
changes from $-\epsilon^s$ to zero is \textit{not} the distance over which a single polar bond breaks.
It is rather the distance over which the \textit{number} of polar bonds shared between two latex particles changes
from its maximum to zero. This will depend on the precise distribution of polar groups through the interfacial region
of the latex particles. Unfortunately this information is lacking. We therefore assumed that the polar groups are
scattered more-or-less randomly throughout the interfacial region of the latex particle. Changes in the
number of shared polar groups should then occur over the same length scale as changes in chain intermixing,
i.e. $w^s = w^e$. The chain intermixing will typically change on a length scale equal to the radius of gyration
of the polymer backbone. For a closely related $M_w = 2203$ kDa PMMA chain, the radius of gyration is estimated to be about 
37 nm \cite{Peterson}. In the model we use $w^s = w^e = 40$ nm, which is 10\% of the particle bare diameter $\sigma_0$.

Given the above estimates for $\epsilon^s$ and $w^s$, we choose the repulsive energy parameter $a^{rep}$ such that the
attractive well $\varphi^{rep} + \varphi^s$ is more or less symmetric, with a characteristic width of $2w^s$, as shown in Fig.~\ref{fig_pmf}.

At this point the potential of mean force of our model is almost fully specified, except for the parameters of the finitely extensible spring,
which determines the internal deformability of a latex particle. 

The range of maximum extension is set to $R_0 = 1000$ nm. Changing the value of $R_0$ hardly has any influence on the
linear rheology, but it can change the nonlinear rheology. For the purpose of this paper,
it is important that $R_0$ is considerably larger than the particle bare diameter $\sigma_0$ so that the particle can accommodate
a large stretch under the influence of strong nonlinear flow. The exact influence of $R_0$ on the nonlinear rheology is a topic of a future paper.

Next we must choose the spring constant $k$. When our model material is under a stress, it can relax this stress by changing
interparticle distances and/or particle extensions. If the force associated with changing a particle extension over a certain distance
is much weaker than the force associated with changing the interparticle distance over the same distance,
the system will always respond by elongating the particles. In the opposite case, the system will always
respond by changing interparticle distances, while keeping the particles effectively spherical (elongation zero).
In real PSA materials these two effects are more or less balanced, so it is important to choose the value of the spring 
constant $k$ roughly equal to the curvature of the total interparticle pair interaction
near its minimum. This way we arrive at $k = 3\ 10^{-3}$ N/m. 

The friction per particle pair was chosen large enough
to suppress Brownian motion (which is also suppressed in the real material) and allow for a large integration time step of $3\ 10^{-5}$ s,
but not so large that interparticle distances and extensions are effectively frozen on the time scales of interest (10$^{-3}$ to 10 s).
A good compromise was found at $\xi_e = 2\ 10^{-4}$ kg/s.

Now we need to specify further the parameters for the transient forces.
Consider again the case where the distance between a pair of neighbouring latex particles is suddenly increased.
Even if the polar groups would be very weak and break easily, the intermixed chains in the interfacial region between the two
latex particles would need a finite time to relax to new configurations. Typically, such a relaxation takes place on the Rouse time
scale, which for our acrylic chain has been measured to be of the order of a millisecond \cite{Lalaso2011}.
We therefore choose $\tau^e = 10^{-3}$ s. Of course the polar
groups also need a finite time to break, and in view of the large binding energy associated with a polar bond, it comes as no
surprise that this time should be relatively large. Unfortunately no measurement or estimates are available, so currently
we treat this as a fit parameter. For our reference material we find good agreement for the dependence of the loss modulus
on frequency if we choose $\tau^s = 1$ s, which seems reasonable.

The transient penalties $\alpha^s$ and $\alpha^e$ have been used to semi-quantitatively tune the simulated storage and loss modulus
to the experimentally measured ones in the range $\omega=10^{-1}$ rad/s to 10$^3$ rad/s. We found that, given the relative smallness of $\tau^e$,
the parameter $\alpha^e$ only influenced the linear rheology at $\omega \ge 10^2$ rad/s. Therefore, $\alpha^s$ was varied first to tune the
linear rheology for low frequencies $\omega < 10^2$ rad/s, leading to $\alpha^s = 10^6 k_BT$, after which $\alpha^e$ was varied to also tune the
high frequencies, leading to $\alpha^e = 10^5 k_BT$.

In a forthcoming paper we will further investigate the influence of each of the model's parameters on the rheology.
Here we already note that the parameters which are expected to be most strongly dependent on the 
PSA chosen for study are $\epsilon^s$, $w^s$, and $\alpha^s$, together characterizing
the distribution and strength of the sticker groups, and $w^e$, $\alpha^e$ and $R_0$, together characterizing chain entanglement
and stretch.










\section{Experimental methods\label{sec_exp}}

Disks of 8 mm diameter were cut out of the dried films by a precise punching tool for the shear flow tests, while for the extensional flow
 characterization, rectangular sized films of 1-2 mm width and about 15 mm length were prepared using a razor blade.

For the linear viscoelastic characterization, small-amplitude oscillatory shear-flow tests were performed with a strain-controlled
rheometer (ARES, TA Instruments) using parallel-plate tools with diameters of 8 mm at frequencies from 10$^{-2}$ rad/s to 10$^{2}$ rad/s
and at temperatures from -45$^{\circ}$C to 90$^{\circ}$C. These results were used (through time-temperature superposition) to extend the
range of accessible frequencies at the reference temperature of 30$^{\circ}$C.

The nonlinear-viscoelastic behavior of the PSA was measured in uniaxial elongational flow using the extensional viscosity fixture
(EVF, TA instruments) on the same rheometer at a temperature of 30$^{\circ}$C and at strain rates of 1 s$^{-1}$ and 10 s$^{-1}$. 
Elongational stress relaxation was measured by cessation of extensional flow at strain rates of 1 s$^{-1}$ for different Hencky strains.

\section{Linear rheology\label{sec_linear}}

We start by exploring up to what point our model PSA behaves linearly. The material is subjected to an oscillatory shear, with the
flow in the $x$-direction, a gradient in the $y$-direction, and a strain given by
\begin{equation}
\gamma(t) = \gamma_0 \sin \left( \omega t \right), \label{eq_oscstrain}
\end{equation}
where $\gamma_0$ is the strain amplitude and $\omega$ the oscillation frequency. In the linear regime, the shear stress response is related to the storage modulus $G'(\omega)$ and loss modulus $G''(\omega)$ according to
\begin{equation}
S_{xy}(t) = \gamma_0 G'(\omega) \sin \left( \omega t \right) + \gamma_0 G''(\omega) \cos \left( \omega t \right).
\end{equation}
Outside the linear regime, the stress response is no longer linear in $\gamma_0$ and possibly
higher harmonics ($3\omega, 5 \omega, \ldots$) will appear.
In all cases, the base frequency responses $G'(\omega)$ and $G''(\omega)$ have been calculated from
\begin{eqnarray}
G'(\omega) &=& \frac{2}{\gamma_0 \tau} \int_{t^*}^{t^*+\tau} S_{xy}(t) \sin(\omega t) \mathrm{d}t, \\
G''(\omega) &=& \frac{2}{\gamma_0 \tau} \int_{t^*}^{t^*+\tau} S_{xy}(t) \cos(\omega t) \mathrm{d}t,
\end{eqnarray}
where $\tau$ is an integer number of oscillation periods ($\tau = 2\pi n/\omega$, with $n$ an integer), and the starting time $t^*$ 
was chosen sufficiently large to avoid any startup effects.

\begin{figure}[tbp]
\begin{center}
\includegraphics[scale=0.5]{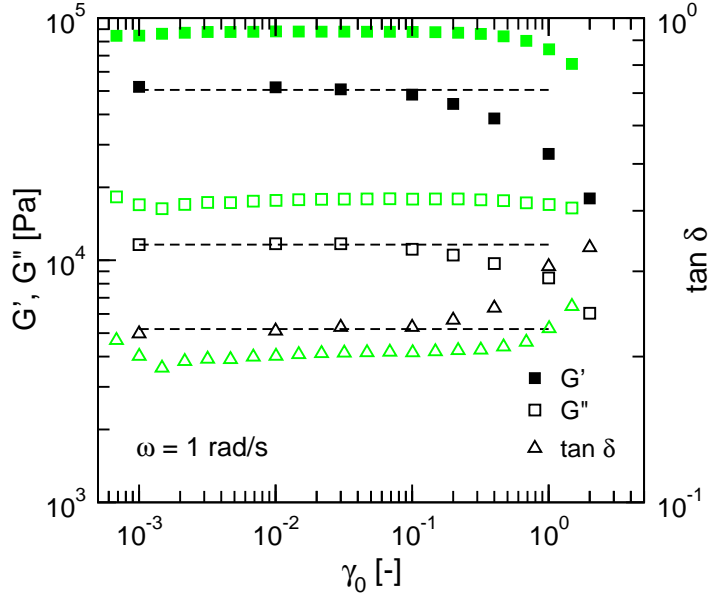}
\end{center}
\caption{\label{fig_G1G2strain}
(color online)
Strain amplitude sweep test of the linear regime for the modeled PSA under oscillatory shear at $\omega = 1$ rad/s.
Shown are the storage modulus $G'$ and loss modulus $G''$ (closed and open squares, left scale)
and the loss tangent $\tan \delta$ (triangles, right scale)
The material behaves linearly (constant $G'$, $G''$ and $\tan \delta$, dashed lines) up to a strain of approximately 0.1.
Experimental results, included for reference (green symbols), are linear up to a strain of approximately 0.4.
}
\end{figure}
For strain amplitudes up to approximately 0.1 our model behaves linearly, as evidenced by the constancy of
$G'$ and $G''$ as a function of $\gamma_0$ in Fig.~\ref{fig_G1G2strain} (black symbols)
and the absence of higher harmonics in the stress response (not shown).
Beyond this value higher harmonics emerge and both the storage and loss modulus start to drop, but the storage modulus drops faster.
As a consequence the loss tangent $\tan \delta = G''/G'$ increases slightly beyond a strain amplitude 0.1.
Experimentally, the linear regime extends up to a strain amplitude of 0.4 after which similar
nonlinear deviations are observed, as shown by the green symbols in Fig.~\ref{fig_G1G2strain}.

In our simulations, the linear storage and loss modulus have been determined for frequencies ranging from 10$^{-1}$ rad/s to 10$^3$ rad/s.
\begin{figure}[tbp]
\begin{center}
\includegraphics[scale=0.5]{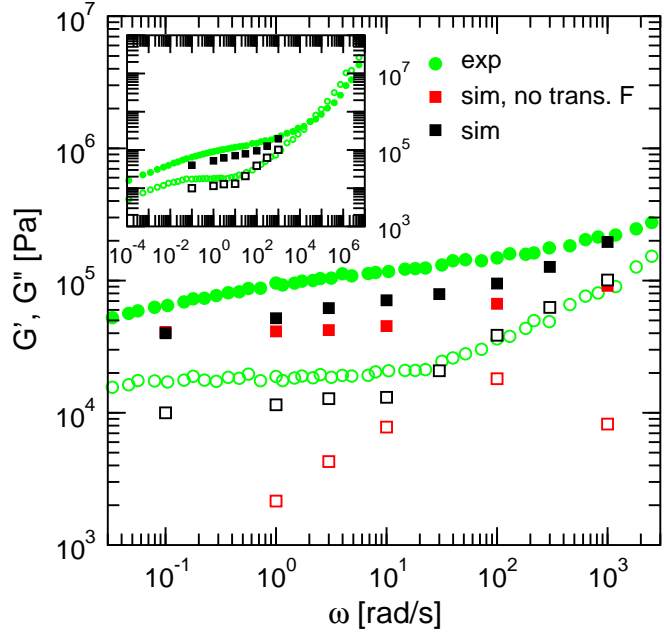}
\end{center}
\caption{\label{fig_G1G2omega}
(color online)
Linear storage (filled symbols) and loss (open symbols) modulus. The main figure shows the results of simulations with (black squares)
and without (red squares) transient forces, as well as experimental results of the reference system (green circles).
The inset shows where our targeted range of frequencies fits in the overall range of experimental frequencies, as obtained by 
time-temperature superposition for a reference temperature of 303 K.
}
\end{figure}
In this range of frequencies the storage modulus $G'$ is consistently larger than the loss modulus $G''$,
hence the material's response is mainly elastic, see the main plot of Fig.~\ref{fig_G1G2omega}.
This is the case both with and without transient forces. Without
transient forces, however, the dependence of the loss modulus on frequency is rather strong and Maxwell-like (open red squares
in Fig.~\ref{fig_G1G2omega}).
The influence of the transient forces, with the current parameters settings, is to increase the (elastic) storage modulus
at higher frequencies and, more significantly, to increase the (dissipative) loss modulus at both high and low frequencies (open black squares).
As a consequence both the storage and loss modulus increase monotonically with increasing
frequency. The increase is slow, spanning less than a decade compared to four decades increase in frequency.
The increase is particularly slow for the loss modulus $G''$ for frequencies up to 10 rad/s.

Fig.~\ref{fig_G1G2omega} (main plot) shows that qualitatively all the above features in the storage and loss modulus are also present in the
experimental system. 
Given the crude way in which we have constructed our mesoscale model, the near-quantitative agreement
with the experimentally determined storage and loss modulus is encouraging.
We expect that only moderate refinements to the model and its parameters will be needed to achieve a full quantitative agreement.
Such refinements will be different for each particular pressure sensitive adhesive; the dependence of the rheology on the
simulation parameters will be explored in a future paper.

For reference, in the inset of Fig.~\ref{fig_G1G2omega} we show the linear rheology of the system in a much
wider range of frequencies, $10^{-4}$ rad/s $< \omega < 10^7$ rad/s, as can be obtained experimentally by time-temperature
superposition. Because our particles are coarse-grained, we do not expect nor do we aim for agreement at frequencies higher
than $10^4$ rad/s. On the other side of the spectrum, because of computational limitations, our simulations are currently limited
to a total run time of order 10 seconds. We can therefore not yet access the $\omega < 0.1$ rad/s regime. With further increase
in speed of computer hardware, combined with parallellization and optimization
of the code, we expect to be able to reach one or two orders of magnitude lower frequencies in the near future. 

\section{Nonlinear extensional rheology\label{sec_nonlinear}}

We will now test the predictions of the mesoscale model in nonlinear flow. One of the most extreme forms of nonlinear flow
is extensional flow. Unlike shear flow, extensional flow has no rotational component, leading to a continual relentless extension
of the particles without a possibility to escape through reorientation.

\begin{figure}[tbp]
\begin{center}
\includegraphics[scale=0.5]{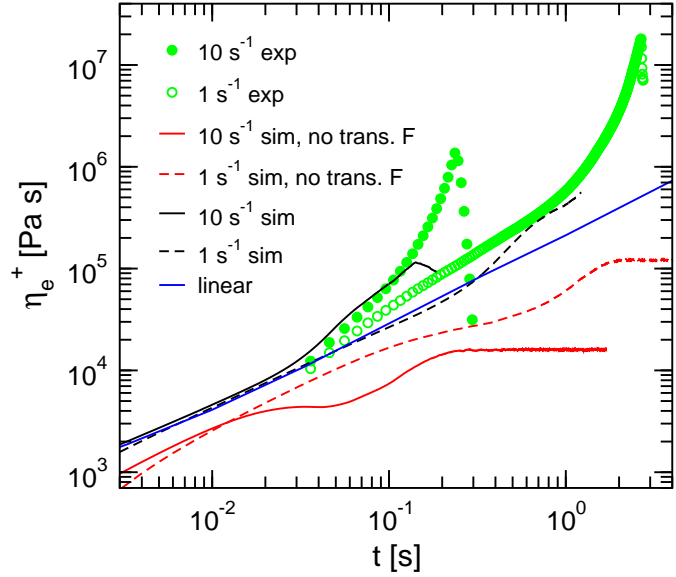}
\end{center}
\caption{\label{fig_extensionvisc}
(color online)
Extensional viscosity growth $\eta^+_e(t)$ as a function of time $t$ after a sudden onset of a planar extensional flow, for two different extension rates $\dot{\epsilon}$ (see legend). We show results for simulated systems with (black lines) and without (red lines) transient forces.
Extensional viscosity growth predicted from the linear rheology (see main text for an explanation) is shown as a blue line.
In the experiments a uniaxial extensional flow is applied; the results for the extensional viscosity growth are shown
for reference (green symbols).
}
\end{figure}

As described in section \ref{sec_flowsim}, we apply a planar extensional flow with constant extension rate $\dot{\epsilon}$
to a previously equilibrated sample.
This flow is suddenly started at $t=0$ and the time-dependent extensional stress $N_1(t) = S_{xx}(t) - S_{yy}(t)$ is measured.
From this we calculate the time-dependent extensional viscosity, defined as
\begin{equation}
\eta^+_e(t) = \frac{N_1(t)}{\dot{\epsilon}}.
\end{equation}
In Fig.~\ref{fig_extensionvisc} we show our results for two extension rates, $\dot{\epsilon} = 10$ s$^{-1}$ (black solid line) and 1 s$^{-1}$
(black dashed line). In the following we will compare these results with (i) a prediction based on the linear rheology,
(ii) results from simulations without transient forces, and (iii) experiments.

For Newtonian liquids the extensional viscosity is related to the shear viscosity through a multiplicative constant,
referred to as the Trouton ratio (Tr). For planar extensional flow Tr$=4$, while for uniaxial extensional flow Tr$=3$.
Similarly, for non-Newtonian liquids the time-dependent extensional viscosity may be compared with the following linear prediction:
\begin{equation}
\eta^{+}_{e,lin}(t) = \mathrm{Tr} \left| \eta^*(\omega)\right|_{\omega=1/t},
\end{equation}
where $\eta^*(\omega)$ is the complex shear viscosity as obtained from linear oscillatory shear measurements.
This prediction, shown as a blue line in Fig.~\ref{fig_extensionvisc}, is in good agreement with the measured extensional viscosity
at small times. At larger times, however, nonlinear effects start to dominate and the extensional viscosity grows faster
than predicted from the linear model. For both extension rates the strain hardening starts at a Hencky strain $\epsilon_H = \dot{\epsilon}t$
of approximately 0.3.

There are clear qualitative differences when comparing simulations with and without transient forces.
In going from lower to higher extension rate, the order of the magnitudes of the viscosity are reversed: without transient forces
(red curves), higher extension rates result in lower extensional viscosities at any particular time beyond the linear regime. Thus without
transient forces, the simulations predict strain thinning behaviour. We can conclude that for this material, the occurrence of strain hardening
behaviour depends crucially on the inclusion of transient forces.

Currently no experimental data is available on \textit{planar} extensional flow of the PSA reference material. However, we can make a 
qualitative comparison with the extensional viscosity measured experimentally in \textit{uniaxial} extensional flow.
These results are shown as green symbols in Fig.~\ref{fig_extensionvisc}. It is encouraging to observe that a similar strain hardening
occurs experimentally, although the onset is somewhat later, at a Henky strain of approximately 1. We expect a
similar onset of nonlinear effects in experimental planar extensional flow. The approximately
threefold increase in the linear strain regime is consistent with the observed threefold increase in the linear regime under oscillatory shear
described in section \ref{sec_linear}. 

\begin{figure}[tbp]
\begin{center}
\includegraphics[scale=0.5]{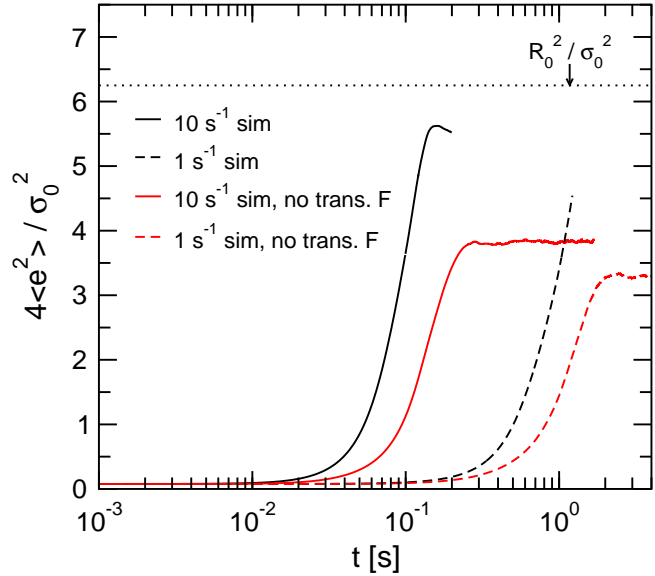}
\end{center}
\caption{\label{fig_extension}
(color online)
Mean square extension $\left\langle (2e)^2 \right\rangle$ of the latex particles as a function of time $t$ after a sudden onset of a planar extensional flow, for two different extension rates $\dot{\epsilon}$ (see legend). The results are normalized by the square of the
unperturbed particle diameter, $\sigma_0^2$.  We show results for simulated systems with (black lines) and without (red lines) transient forces.
The maximum extensibility used in this paper, $R_0 = 2.5 \sigma_0$, is indicated by the horizontal dotted line.}
\end{figure}

At a certain Henky strain a maximum extensional viscosity is reached. Experimentally,
this occurs at a higher Henky strain than in our simulations. This could be related to the fact that the maximum extension of the latex
particles in the simulations is lower than in reality, leading to an extensional stress which reaches its maximum sooner.
We test this hypothesis in Fig.~\ref{fig_extension}, which shows that indeed the mean square extension is close to the maximum extension
$R_0^2$ at the time when the extensional viscosity reaches its maximum. 
The real polymeric particles can be extended to a larger amount, leading to larger stresses and larger maximum strain values.
In future work we will vary the value of $R_0$ and investigate its influence on the nonlinear rheology. 

\begin{figure}[tbp]
\begin{center}
\includegraphics[scale=0.5]{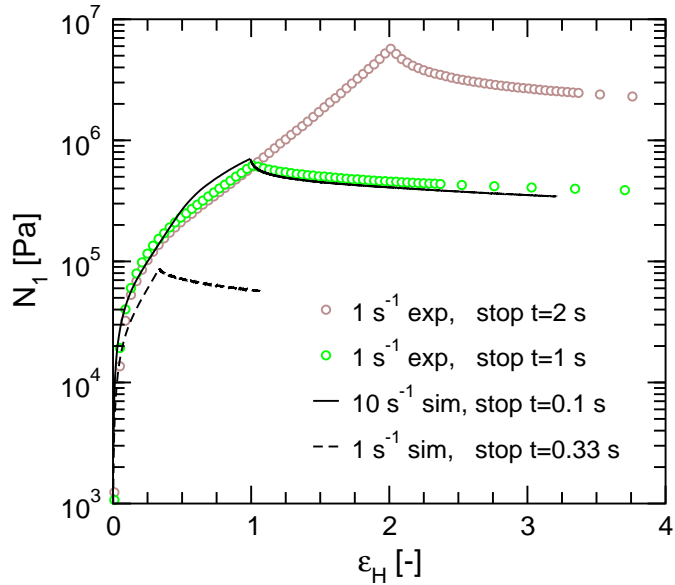}
\end{center}
\caption{\label{fig_extensionrelax}
(color online)
Relaxation of extensional stress after a sudden stop of planar extensional flow, for an initial extension rate of $\dot{\epsilon} = 1$ s$^{-1}$ stopped at $t=0.3$ s (dashed line) and an initial extension rate of $\dot{\epsilon} = 10$ s$^{-1}$ stopped at $t=0.1$ s (solid line).
Results are shown as a function of Hencky strain $\epsilon_H = \dot{\epsilon}t$ as calculated from the initial extension rate.
In the experiments a uniaxial extensional flow is applied; results are shown for the extensional stress relaxation following an initial extension
rate of $\dot{\epsilon} = 1$ s$^{-1}$ which is stopped at $t=1$ s or $t=2$ s (green and brown circles, respectively).
}
\end{figure}

Finally, we study the relaxation of extensional stress after a sudden stop of extensional flow. 
This is a sensitive probe of the relaxation processes that take place as the system evolves from a highly
non-equilibrium to a new equilibrium state. Figure~\ref{fig_extensionrelax} shows the
stress when a sample is (planar) extended with an extension rate $\dot{\epsilon} = 1$ s$^{-1}$, which is suddenly stopped at a Hencky strain of 0.33 (black dashed line), as well as the stress when a sample is extended with an extension rate $\dot{\epsilon} = 10$ s$^{-1}$, which is suddenly stopped at a Hencky strain of 1.0 (dashed black line).
We observe that the extensional stress has a small initial relaxation, which is faster when the Hencky strain is higher. 
Soon hereafter we encounter an extremely slow extensional stress relaxation.

Using our experimental equipment, it is difficult to measure extensional stress relaxation after a high initial extension rate of 
$\dot{\epsilon} = 10$ s$^{-1}$. It is however possible to measure it after extending with a rate of 1 s$^{-1}$.
Figure~\ref{fig_extensionrelax} shows these results for two experiments, in which uniaxial extensional flow is stopped after a Hencky strain
of 1 (green circles) or 2 (brown circles), respectively. The experiments confirm the simulation predictions of a fast initial relaxation which is
faster for higher Hencky strain. Furthermore, the rate of the very slow stress relaxation that follows after the initial fast decay is very similar
to what our simulations have predicted. 

\section{Conclusion and outlook\label{sec_concl}}

We have shown that the rheology of pressure sensitive adhesives can be reproduced nearly quantitatively
by a coarse-grained model if, besides conservative forces, also transient forces between the constituent latex particles are
taken into account. The conservative forces consist of attractive interactions caused by polar sticker groups,
a soft repulsion and a nonlinear elastic spring associated with deformations of the latex particle.
The transient forces arise from temporary deviations in the number of stickers and chain intermixing between pairs of latex particles.

Our model reproduces nearly quantitatively the dominance of the storage over the loss modulus, their slow and monotonous increase
with frequency, and predicts nonlinear elongational effects such as the onset of extensional strain hardening and the characteristics of the 
relaxation of the extensional stress after cessation of extensional flow.

Further improvements of the model and tweaks of the parameters are possible.
For example, the relaxation times $\tau^s$ and $\tau^e$ are now taken
as constants, but may actually depend on the distance between a pair of particles.
This will effectively generate an even wider spectrum of relaxation times.
Also the friction on the centre-of-mass coordinate of each particle is now taken to be the same
as the friction on the extension coordinate of a particle, and both are based on the amount of intermixing (overlap) with neighbouring particles. 
These frictions may be (slightly) different, and may also depend on the number of stickers shared with neighbouring particles.
Finally, the value of the maximum extension of a latex particle is important in determining the strain (both in shear and extension) at
which nonlinear effects in the rheology start to play a role. 
We expect that such tweaks to the model and its parameters will bring the rheological predictions in quantitative agreement 
with experimental results.

The purpose of this paper was to introduce the mesoscale model and to show the importance of including transient forces.
The qualitative and near-quantitative agreement with experimental results motivates us (i) to further investigate the influence of
each of the model's parameters on the rheology, (ii) to subject the model to more complex types of flow, and (iii) to study (onset of)
flow-induced instabilities such as cavity formation. Work along these lines is currently in progress.

\section*{Acknowledgments} We thank Evelyne van Ruymbeke and Roland Keunings for stimulating and insightful discussions.
We are indebted to the CISM supercomputing centre for providing us with computational facilities.
This work is part of NMP ``MODIFY'', funded by the European Union (Framework Programme 7).

\newpage

\end{document}